\documentclass{mem}
\usepackage{natbib}\usepackage{txfonts}\usepackage{balance}
\usepackage{graphicx}
\idline{75}{282}
\begin{document}

\def\teff{$T\rm_{eff }$}
\def\kms{$\mathrm {km s}^{-1}$}

\newcommand{\DontShow}[1] {}
\newcommand{\ToDo}[1] {\textbf{ToDo: #1}}
\def\tens#1{\bar{\bar{\ensuremath{\mathsf{#1}}}}}
\newcommand{\papa}[2] { \frac{\partial #1}{\partial #2} }
\newcommand{\papaco}[3] {{ \left. \frac{\partial #1}{\partial #2}\right|_{#3}}}
\newcommand{\tildepapaco}[3] {{ \left. \widetilde{\frac{\partial #1}{\partial #2}}\right|_{#3}}}
\newcommand{\COBOLD}{CO5BOLD}
\newcommand{\I}[1] {i{\rm #1}}
\newcommand{\m}[1] {m{\rm #1}}
\newcommand{\n}[1] {n{\rm #1}}
\newcommand{\mghost}[1] {m{\rm #1}_{\rm ghost}}
\newcommand{\nghost}[1] {n{\rm #1}_{\rm ghost}}
\newcommand{\dx}[1] {\Delta x{\rm_#1}}
\newcommand{\xc}[1] {x_{\mathrm{c} #1}}
\newcommand{\xb}[1] {x_{\mathrm{b} #1}}
\newcommand{\g}[1]{g{\rm_#1}}
\newcommand{\phic}[1]{\Phi_{\mathrm{c} ,\,  #1}}
\newcommand{\phib}[1]{\Phi_{\mathrm{b} ,\,  #1}}
\newcommand{\V}[1]{v_{#1}}
\newcommand{\B}[1]{B_{#1}}
\newcommand{\tildeV}[1]{\tilde{v}{\rm #1}}
\newcommand{\Press} {P}
\newcommand{\ei} {e_\mathrm{i}}
\newcommand{\eg} {e{\rm g}}
\newcommand{\eikb} {e{\rm ikb}}
\newcommand{\eikbg} {e{\rm ikbg}}
\newcommand{\etot} {e_\mathrm{t}}
\newcommand{\tildeei} {\tilde{e}{\rm i}}
\newcommand{\tildeeik} {\tilde{e}{\rm ik}}
\newcommand{\tildeek} {\tilde{e}{\rm k}}
\newcommand{\tildeeip} {\tilde{e}{\rm ip}}
\newcommand{\eikp} {e{\rm ikp}}
\newcommand{\tildeeikp} {\tilde{e}{\rm ikp}}
\newcommand{\dPdei} {\papaco{P}{\ei}{\rho}}
\newcommand{\dPdrho} {\papaco{P}{\rho}{\ei}}
\newcommand{\drhoeidP} {\papaco{\rhoei}{P}{s}}
\newcommand{\drhoeidrho} {\papaco{\rhoei}{\rho}{P}}
\newcommand{\tildedrhoeidP} {\tildepapaco{\rhoei}{P}{s}}
\newcommand{\rhov}[1]{\rho v{\rm_#1}}
\newcommand{\rhoei} {\rho\ei}
\newcommand{\rhoeik} {\rho e{\rm ik}}
\newcommand{\rhoeip} {\rho e{\rm ip}}
\newcommand{\rhoek} {\rho e{\rm k}}
\newcommand{\rhoeg} {\rho e{\rm g}}
\newcommand{\rhoeikg} {\rho e{\rm ikg}}
\newcommand{\CCour} {C_{\rm Courant}}
\newcommand{\CCourmax} {C_{\rm Courant,max}}
\newcommand{\iup} {i_{\rm up}}
\newcommand{\cs} {c_\mathrm{s}}
\newcommand{\vc} {v_\mathrm{c}}
\newcommand{\wL} {w{\rm L}}
\newcommand{\wR} {w{\rm R}}
\newcommand{\Pred}{{P_{\rm st}}}
\newcommand{\Deltat}{\mbox{\small $\Delta$} t}
\newcommand{\dvdx}[2] { \frac{\Delta \V{#1}}{\dx{#2}} }
\newcommand{\Source}{S}
\newcommand{\dSource}{\Delta \Source}

\newcommand{\Int}{I}
\newcommand{\dInt}{\Delta \Int}
\newcommand{\ImS}{\hat{I}}
\newcommand{\dImS}{\Delta \hat{I}}
\newcommand{\Ired}{\tilde{I}}
\newcommand{\dIred}{\Delta \tilde{I}}

\newcommand{\Dx}{\Delta x}

\newcommand{\dtau}{\Delta \tau}
\newcommand{\dedtau}[1] { \frac{{\rm d} #1}{{\rm d} \tau} }
\newcommand{\dedtaunu}[1] { \frac{{\rm d} #1}{{\rm d} \tau_{\nu}} }
\newcommand{\dededtau}[1] { \frac{{\rm d}^2 #1}{{\rm d} \tau^2} }
\newcommand{\dededtaunu}[1] { \frac{{\rm d}^2 #1}{{\rm d} \tau_{\nu}^2} }

\newcommand{\dedx}[1] { \frac{{\rm d} #1}{{\rm d} x} }

\newcommand{\elt} {\widehat{e}}

\newcommand{\wII}{w(2)}
\newcommand{\wIII}{w(3)}

\newcommand{\Snu}{\ensuremath{S_\nu}}
\newcommand{\Jnu}{\ensuremath{J_\nu}}
\newcommand{\taunu}{\ensuremath{\tau_\nu}}

\bibliographystyle{aa}

\title{Atmospheres from very low-mass stars to extrasolar planets}
 
 \subtitle{}

\author{
F. \, Allard \& D. Homeier
          }

  \offprints{F. Allard}

\institute{
CRAL, UMR 5574, CNRS,
Universit\'e de Lyon, 
\'Ecole Normale Sup\'erieure de Lyon, 
46 All\'ee d'Italie, F-69364
Lyon Cedex 07, France
\email{fallard@ens-lyon.fr}
}

\authorrunning{Allard }

\titlerunning{modeling the stellar-substellar transition}

\abstract{
Within the next few years, several instruments aiming at imaging extrasolar planets 
will see first light. In parallel, low mass planets are being searched around red dwarfs 
which offer more favorable conditions, both for radial velocity detection and transit studies, 
than solar-type stars.  We review recent advancements in modeling the stellar to substellar 
transition. The revised solar oxygen abundances and cloud models allow to reproduce the 
photometric and spectroscopic properties of this transition to a degree never achieved before,
but problems remain in the important M-L transition characteristic of the \teff\ range of 
characterisable exoplanets.  

}
\maketitle{}

\section{Introduction}

Since spectroscopic observations of very low mass stars (late 80s), brown dwarfs (mid 90s), and extrasolar planets
(mid 2000s) are available, one of the most important challenges in modeling their atmospheres and spectroscopic 
properties lies in high temperature molecular opacities and cloud formation. K dwarfs show the onset of formation 
metal hydrides (starting around \teff\ $\sim 4500$\,K), TiO and CO (below \teff\ $\sim 4000$\,K), while water vapor 
forms in early M dwarfs (\teff\ $\sim 3900-2000$\,K), and methane, ammonia and carbon dioxide are detected in 
late-type brown dwarfs (\teff\ $\sim 300-1600$\,K) and in extrasolar giant planets.  Cloud formation is also an 
important factor in the detectability of biosignatures, and for the habitability of exoplanets 
\citep[Kasting 2001]{Paillet2005}\nocite{Kasting2001}.  

Extrasolar planets for which we can currently characterize their atmospheres are either those observed by transit 
(\teff\ $\sim 1000-2000$\,K depending on their radius relative to that of the central star) or by imaging (young planets 
of \teff\ $\sim 500-2000$\,K depending on their mass and age). 
Several infrared integral field spectrographs combined with coronagraph and adaptive optic instruments are
coming online before 2013 (SPHERE at the VLT, the Gemini Planet Imager at Gemini south, Project1640 at Mount Palomar, etc.). 
The E-ELT 41\,m telescope in Spain due around 2020 will also be ideally suited for planet imaging.  

M dwarfs are the most numerous stars, constituting 80\% of the stellar budget of the Galaxy, and around 600 brown dwarfs and planets 
are currently known despite their faintness in the solar neighborhood vicinity.  Single very low mass (VLM) stars and brown dwarfs are 
therefore more directly observable and characterizable then exoplanets. They represent, beyond their own importance, a wonderful  
testbed for the understanding of exo-planetary atmospheric properties together with solar system studies. Planets can even share the 
atmospheric composition of brown dwarfs of same \teff\ (see section \ref{s:Comp} below). 

The models developed for VLMs and brown dwarfs are therefore a unique tool, if they can explain the stellar-substellar transition, for 
the characterization of imaged exoplanets.  In this paper, we review the ability of recently published models in reproducing constraints 
along the M-L-T spectral transition.

\vspace{-2.6cm}
\section{Model Construction}
The modeling of the atmospheres of VLMs has evolved (as here illustrated with the 
development of the \texttt{PHOENIX} atmosphere code) with the extension of computing capacities from an analytical 
treatment of the transfer equation using moments of the radiation field \cite[]{AllardPhDT90}, to a line-by-line opacity 
sampling in spherical symmetry \cite[1997 and Hauschildt et al. 1999]{Allard1994}\nocite{ARAA97,NGa} and more 
recently to 3D radiation transfer \cite[]{SHB2010}. In parallel to detailed radiative transfer in an assumed static environment, 
hydrodynamical simulations have been developed to reach a realistic representation of the granulation and its 
induced line shifts for the sun and sun-like stars \citep[see e.g. the review by][]{Freytag2012} by using a non-grey
(multi-group binning of opacities) radiative transfer with a pure blackbody source function (scattering is neglected).

To illustrate the various assumptions made by constructing model atmospheres, let us begin with the description of the 
equations of ideal magneto\-hydro\-dynamics (MHD) --- adapted here for the stellar case by specifying the role of gravity, 
radiative transfer, and energy transport --- which are themselves a special case (no resistivity) of the more 
general equations \cite[see for example ][]{Landau1960}.  These are written in the compact vector notation as:

\begin{equation}
\label{eq:MHD-equations}\setlength{\arraycolsep}{2pt}
\begin{array}{cllll}
  \displaystyle\!\papa{\rho}{t}
  & \!+\! &\! \mathbf{\nabla}\!\cdot\!(\rho\mathbf v)                                \!&\! = & 0         \enspace,      \\[2ex]
  \displaystyle\!\papa{\rho\mathbf{v}}{t} 
  & \!+\! &\! \mathbf{\nabla}{\cdot} (\rho\mathbf{vv} +
        (P+\displaystyle{\scriptsize\frac{1}{2}}\mathbf{B}\!\cdot\!\mathbf{B}) \mathbf{I} -
                               \mathbf{BB})                                      \!&\! = \!&\! \rho\mathbf{g},          \\[2ex]
  \displaystyle\!\papa{\mathbf{B}}{t}
  & \!+\! &\! \mathbf{\nabla}\!\cdot\!(\mathbf{vB} - \mathbf{Bv})                     \!&\! = \!&\! 0    \enspace, \\[2ex]
  \displaystyle\!\papa{\rho\etot}{t}
  & \!\!+\! &\! \mathbf{\nabla}\!\cdot ( (\rho\etot+P+
        \displaystyle {\scriptsize\frac{1}{2}} \mathbf{B}\!\cdot\!\mathbf{B})\mathbf{v}   & & \\[2ex]
   &&  ~~~~~~~~~~~~~~~ - (\mathbf{v}\!\cdot\!\mathbf{B})\mathbf{B} + \mathbf{F_{\rm rad}}) \!&\! = \!&\! 0      \enspace.       
\end{array}
\end{equation}

The vectors are noted with boldface characters, while scalars are not. For example, $P$ is the gas pressure, 
$\mathbf{\rho}$ the mass density, $\mathbf{g}$ the gravity, and $\mathbf{v}$ is the gas velocity at each point in space.  
$\mathbf{B}$ is the magnetic field vector, where the units were chosen such that the magnetic permeability 
$\mu$ is equal to one.
$\mathbf{I}$ is the identity matrix and
$\mathbf{a}\cdot\mathbf{b} = \sum_k a_k b_k$ the scalar product
of the two vectors $\mathbf{a}$ and $\mathbf{b}$.
The dyadic tensor product of two vectors 
$\mathbf{a}$ and 
$\mathbf{b}$ is the tensor 
$\mathbf{ab} = \mathbf{C}$ with elements
$c_{mn} = a_m b_n$
and the $n$th component of the divergence of the tensor $\mathbf{C}$ is 
$\left(\mathbf{\nabla} \cdot \mathbf{C}\right)_n  = \sum_m \partial c_{mn}/\partial x_m$.
In this case, the total energy is given by
\begin{equation}
\label{eg:TotalEnergy}
\rho\etot=  \rho\ei 
          + \rho\scriptsize\frac{1}{2}\mathbf{v}\cdot\mathbf{v} 
          + \scriptsize\frac{1}{2}\mathbf{B}\cdot\mathbf{B}
          + \rho \Phi \enspace , 
\end{equation}
where $\ei$ is again the internal energy per unit mass, and $\Phi$ the  gravitational potential.
The additional constraint for the absence of magnetic monopoles,
\begin{equation}
\label{eq:NoMonopoles}
\mathbf{\nabla}\cdot \mathbf{B}=0 \enspace ,
\end{equation}
must also be fulfilled. 

The first, third, and last equations in eq.~\ref{eq:MHD-equations} correspond to the mass, magnetic field, and energy conservation, 
while the second equation is the budget of forces acting on the gas. In the case of stellar astrophysics, gravitational acceleration
is an important source term, while the radiative flux participates in the energy budget. Further assumptions are made 
in the numerical solution of these equations to address different astrophysical problems in very different regimes. 
The chromospheres correspond to a regime of high Mach numbers and strong magnetic fields where ionized gas has to 
follow the magnetic field lines, and where the radiative transfer must be solved for the case of a non-ideal gas.
The photospheric convection simulations correspond to a regime where the thermal and convective turnover 
timescales are comparable i.e. Mach numbers are around 1, and the non-local radiative transfer must be solved, often for an 
ideal gas. And the interior convection and/or dynamo simulations correspond to a regime where the thermal timescale 
is much larger then the turnover timescale, which in turn is much larger then the acoustic timescale.
The radiative flux can be approximated by the diffusion approximation, and the magnetic field lines are dragged by the ionized gas.

Radiation hydrodynamical (RHD) simulations ignore by definition the magnetic field terms in equation \ref{eq:MHD-equations}.
This is a good approximation when modeling the neutral photosphere (where most of the emitted flux emerges) of low mass, 
very low mass stars, and brown dwarfs --- with the exception of the ultraviolet and visual spectral range of flaring stars and for the 
resulting emission lines.  RHD simulations, especially in 2D and 3D, are computationally expensive, and, 
when treating -- if at all -- radiative transfer, can currently be performed only for a restricted number of wavelengths, or wavelength 
bins (typically 4 to 12). In the case of solar-like photospheres, RHD simulations using $140^2 \times 150$ grid points over 5 hours 
of stellar time CO5BOLD required on parallel computers (2 nodes)  several CPU months \citep{Ludwig2009}.  In the case of red 
and brown dwarf simulations, local 2D cases with a resolution of  $400 \times 300$ covering 2 days of stellar time 
CO5BOLD \citep{Freytag2010} required 1 month of CPU time.  

The classical approach for interior and atmosphere models consists in simplifying the problem for a gain of computing efficiency, 
neglecting the magnetic field, convective and/or rotational motions and other multi-dimensional aspects of the problem, and 
assuming that the averaged properties of stars can be approximated by modeling their properties radially (uni-dimensionally) 
and statically. We also assume that the atmosphere does neither create nor destroy the radiation emitted through it. Neglecting 
motions in modeling the photospheres of VLM stars, brown dwarfs, and planets is acceptable since the 
convective velocity fluctuation effects on line broadening is hidden by the strong van der Waals broadening prevailing in these 
atmospheres.   But this is not the case of the impact of the velocity fields on the cloud formation and wind processes (see section 
\ref{s:Mixing} below). In this case, equation \ref{eq:MHD-equations} reduces to the so-called hydrostatic equation and constant 
flux approximation for the radial or $z$ direction used in classical models:

\begin{equation}
\label{eq:classical}\setlength{\arraycolsep}{2pt}
\begin{array}{cllll}
  \displaystyle\papa{P}{r}                     &   &                                                                                      & = & - {\rho}g             \enspace,         \\[2ex]
  \displaystyle\papa{F_{\rm rad}}{r}   &  = &  \displaystyle\papa{(\int \, {F_\lambda} \, \mathrm{d\lambda})}{r}           & = & 0      \enspace.       
\end{array}
\end{equation}

This allows computing the interior evolutive properties of stars throughout the Hertzsprung-Russell diagram, and 
to solve the radiative transfer in the atmosphere for a much larger number of wavelengths (line-by-line or opacity 
sampling) or wavelength bins (Opacity Distribution Function or ODF, K-Coefficient) compared to R(M)HD simulations. 
Classical model atmospheres impose therefore the independent parameter F$_{\rm rad}$ ($= \sigma$ {\teff}, where $\sigma$ 
is the Stefan-Boltzman constant) and compute F$_\lambda$ so that, after model convergence, the target F$_{\rm rad}$ is reached. 
Other independent parameters are the surface gravity $g$ and the abundances of the elements $\epsilon_i$.  
This makes it possible to create extensive databases of synthetic spectra and photometry that provide the basis for the interpretation 
of stellar observations. 

All the model atmospheres compared in this review are classical models in this sense, and differ mainly in the completeness 
and accuracy of their opacity database ()including their cloud model assumptions), and the assumed solar abundances used 
for the particular grid shown. They must resolve the radiative transfer for the entire spectral energy distribution (as can be seen 
from eq. \ref{eq:classical}) with a good enough spectral resolution to account for all cooling and heating processes. 

Classical model atmospheres differ also from one another in their construction philosophy, which is linked to their period of initial 
development. The code by \cite[2002, 2005]{Tsuji1965}\nocite{tsuji02}, the ATLAS code by \cite{ATLAS1973} and \cite{ATLAS92004}, 
and the MARCS code by \cite[2008]{MARCS1975}\nocite{MARCS2008} have seen the punched computer cards and the need to spend all 
efforts in saving characters and computer time. These models pre-tabulate their opacities ($\sum_i\kappa_i({\lambda})= \sum_i n_i ~ 
{\sigma}_i({\lambda})$, where $\kappa_i({\lambda})$  [cm$^{-1}$] and ${\sigma}_i({\lambda})$  [cm$^2$]  are the opacity coefficient and 
cross-section at the wavelength $\lambda$, and $n_i$ [cm$^{-3}$] is the number density of species~$i$ i.e. atoms, molecules or grains) 
to interpolate them later during the model atmosphere execution.  
The \texttt{PHOENIX} code \citep[2012]{Allard1994} on the other hand, also to distinguish itself from its forefathers, took the approach 
of computing the opacities during the model execution (or on-the-fly). This involves computing the opacities for billions of atomic 
and molecular transitions on-the-fly, though with a selection of the most important lines. This different approach makes  \texttt{PHOENIX} 
much slower then former codes, but allows to take into account more consistently important physical phenomena, such as those involving 
a modification of local elemental abundances along the atmospheric structure (e.g. non-LTE, photoionisation, diffusion and cloud formation).

Therefore, and especially in the  \texttt{PHOENIX} case, the computational requirements of classical model atmospheres, even nowadays, 
preclude in practice modeling globally a star from its interior to its photospheric layers. Besides, an eventual classical static 1D model
appears less interesting then global RHD simulations. This is  becoming possible even with rotation but of course at the cost of some severe
approximations at this point:  the innermost core is replaced by an adapted potential function \citep{Sun_Rotation07}.  These RHD simulations 
of main sequence stars, brown dwarfs, and planets have also to be scaled down significantly in radius to preserve the ability to resolve convective 
cells and timescales of important processes such as cloud formation.   An alternative approach used by many authors is therefore to neglect 
small scale phenomena and model only larger scales, such as global circulation around the  planetary surface \citep[see for example][Showman 
et al. 2009, Dobbs-Dixon et al. 2010]{Koskinen2007}\nocite{Showman2009,Dixon2010}. The challenge of such hydrodynamical simulations 
nowadays is to account for all the most important opacities, in particular scattering, in solving the radiative transfer and hydro equations while
keeping the computing time for the model within reasonable limits.


\section{Molecular opacities}

While earlier work has been developed for the study of red giant
stars, the pioneering work on the modeling of VLM atmospheres has been
provided by \cite{Mould75}, \cite{AllardPhDT90} and \cite{PhDTKui91} using
a band model or the Just Overlapping Line Approximation (JOLA) opacities
developed by \cite{Kivel52} and adapted for astrophysical use by
\cite{Golden67}.

More realistic model atmospheres and synthetic spectra for VLMs, brown
dwarfs. and extrasolar planets  using line-by-line or opacity sampling techniques
have been made possible thanks to the development of accurate opacities calculated 
often ab initio for atmospheric layers where temperatures can reach 3000\,K. 
The process of improvements was especially remarkable in the case of 
water vapor line lists.  Indeed, water vapor has seen an important 
evolution through the years from band model approximations to straight 
means based on hot flames experiments, and then to ab initio computations.  
Nevertheless, the atmosphere models have failed to reproduce the 
strength of the water bands that shape the low resolution
($R\le$\,300) infrared spectral energy distributions (hereafter SED) of M dwarfs. 
At the lower temperatures of brown dwarfs, methane and ammonia rival
the effect of water. 
The discrepancies in the model synthetic spectra were therefore believed
to be due to inaccurate or incomplete molecular opacities.  In
particular water vapor was suspected because the 
discrepancies were observed at infrared wavelengths in the relative
brightnesses of the flux peaks between water vapor bands.  

\begin{figure}[!ht]
  \includegraphics[width=6.9cm]{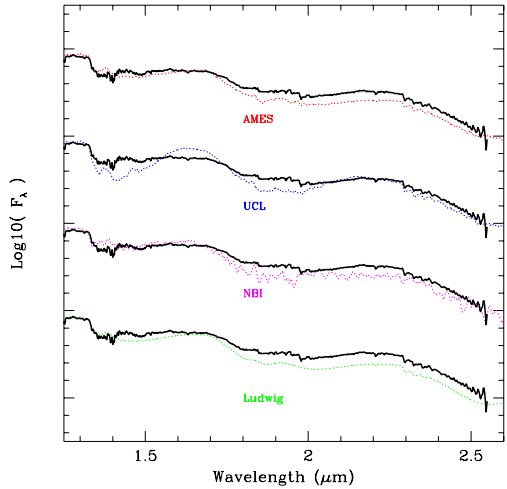}
  \caption{\footnotesize
    Fig. 1 of the review article by Allard et al. (2012). The near-infrared SED of 
    VB10 is compared to synthetic spectra (\teff\ $=2800$\,K, 
    $\mathrm{log}\,g=5.0$, [M/H]$=0.0$, $\Delta\lambda=50$\,{\AA}) from 
    diverse model grids published through the years.  All models (except the 
    NextGen/UCL case)  underestimate the flux in the $K$ bandpass  by 0.1 to 0.2 dex.}
  \label{f:allard_f1_water}
\end{figure}

In Fig.~\ref{f:allard_f1_water} the models are compared to the infrared spectrum of the M8e
dwarf VB10. One can see that the water vapor opacity profile, which shape this part of the 
spectrum, has strongly changed over time with the improvement of computational capacities 
and a better knowledge of the interaction potential surface. 
Only the most recent ab initio results \citep[and the BT2 line list by Barber et al. 2006]{AMESH2O}\nocite{BT2H2O} 
confirm the earliest hot flames laboratory experiment results by \cite{LudwigH2O}. 
Nevertheless, a lack of flux persist in the $K$ bandpass in the models even using 
the most recent BT2 opacity profile \citep[e.g. the BT-Settl models of][]{Allard2012}.  
Only the UCL line list \citep[due to incompleteness, 
and with much of its deviations canceling out over the bandpasses]{UCL94H2O} could produce 
seemingly correct $J-K$ colors, and could allow some success of this so-called NextGen \citep{NGa}
model atmospheres grid in the VLM stellar regime.

In the substellar regime, the composition of brown dwarfs varies rapidly with decreasing {\teff},
and the variation is responsible for the immense change in their SED across the very narrow \teff\ regime
of the M-L-T spectral transition. If water vapor opacities only became recently reliable, this is not 
the case of the more complex methane molecule which is so important in brown dwarfs, and 
planetary atmospheres. The ExoMol Project supported by an ERC to Jonathan Tennyson 
(University College London) will allow important advances on these fronts in the coming years. 
A new ammonia line list is already available through this project \citep{Yurchenko2011}.

\section{Mixing}\label{s:Mixing}

Stars becomes fully convective throughout their interior and convection reaches furthest out in the optically thin regions 
of the photosphere  in M3 and later dwarfs with \teff\ below 3200\,K \citep[Chabrier \& Baraffe 2000]{AllardPhDT90}\nocite{CB00}.  
In most model atmospheres discussed in this review paper, the convective energy transfer is treated using the 
Mixing Length Theory \citep[or MLT, see][]{KW1994}, using at best a unique fixed value of the mixing length of 1.0  
(1.25 for the ATLAS9 models, 1.5 for the MARCS models, etc). 
However, since convection becomes efficient in M dwarfs, the precise value of the mixing length matters only for the deep 
atmospheric structure and as a surface boundary condition for interior models.

\citet{Ludwig2002} and \citet{Ludwig2006} have been able to compare the \texttt{PHOENIX} thermal structure obtained 
using the MLT with that of RHD simulations. They showed that the MLT could reproduce 
adequately (except for the overshoot region) the horizontally averaged thermal structure of the hydro simulations when using  
an adequate value of the mixing length parameter.  This value has been estimated for M dwarfs to vary with surface 
gravity from $\alpha$=$l/H_p$=1.8 to 2.2 (2.5 to 3.0 for the photosphere).  

The BT-Settl models use the mass and surface gravity dependent prescription of  \citet{Ludwig1999}  for hotter 
stars, together with an average (2.0) of the values derived for M dwarfs by \cite[2006]{Ludwig2002}\nocite{Ludwig2006}.  
They use as well the micro-turbulence 
velocities from the radiation hydrodynamical simulations \citep{Freytag2010}, and the velocity field from RHD simulations
from \cite{Ludwig2006} and \cite{Freytag2010} to calibrate the scale height of overshoot, which becomes important in forming 
thick clouds in L dwarfs but is negligible for the SED of VLMs and brown dwarfs otherwise.    

\cite{Freytag2010} have indeed addressed the issue of mixing and diffusion in VLM atmospheres by 2D RHD simulations, using the 
\texttt{PHOENIX} gas opacities in a multi-group opacity scheme, and forsterite with geometric cross-sections. These simulations 
assume efficient nucleation, using initial monomer densities estimated from the total available density of silicon (least abundant 
element in the solar composition involved in forsterite).  They found that gravity waves form at the internal convective-radiative 
boundary, and play a decisive role in cloud formation, while around \teff\ $\approx 2200$\,K the cloud layers become
thick enough to initiate cloud convection, which dominates in the mixing.

\section{Atmospheric composition}\label{s:Comp}

The composition of the atmospheres of stars, brown dwarfs, and planets is a function of \teff\ (radiation either due to internal 
heat from nuclear fusion and contraction or from irradiation by a parent star), of surface gravity to a lesser extent, and of the 
elemental abundances of the initial gas from which the star or stellar system is formed. Stellar model atmospheres assume 
scaled solar abundances for all elements relative to hydrogen. Additionally, some enrichment of  $\alpha$-process elements  
(C, O, Ne, Mg, Si, S,  Ar, Ca, and Ti) resulting from a "pollution" of the star-forming gas by the explosion of a supernova 
is appropriate in the case of metal-poor subdwarfs of the Galactic thick disk, halo, and globular clusters, and the stars in 
the high stellar density environment  towards the galactic center \citep{Gaidos2009}.  

\subsection{The revision of solar abundances}

\begin{figure*}[ht!]
\hspace{-1.3cm}
 \includegraphics[width=8.2cm]{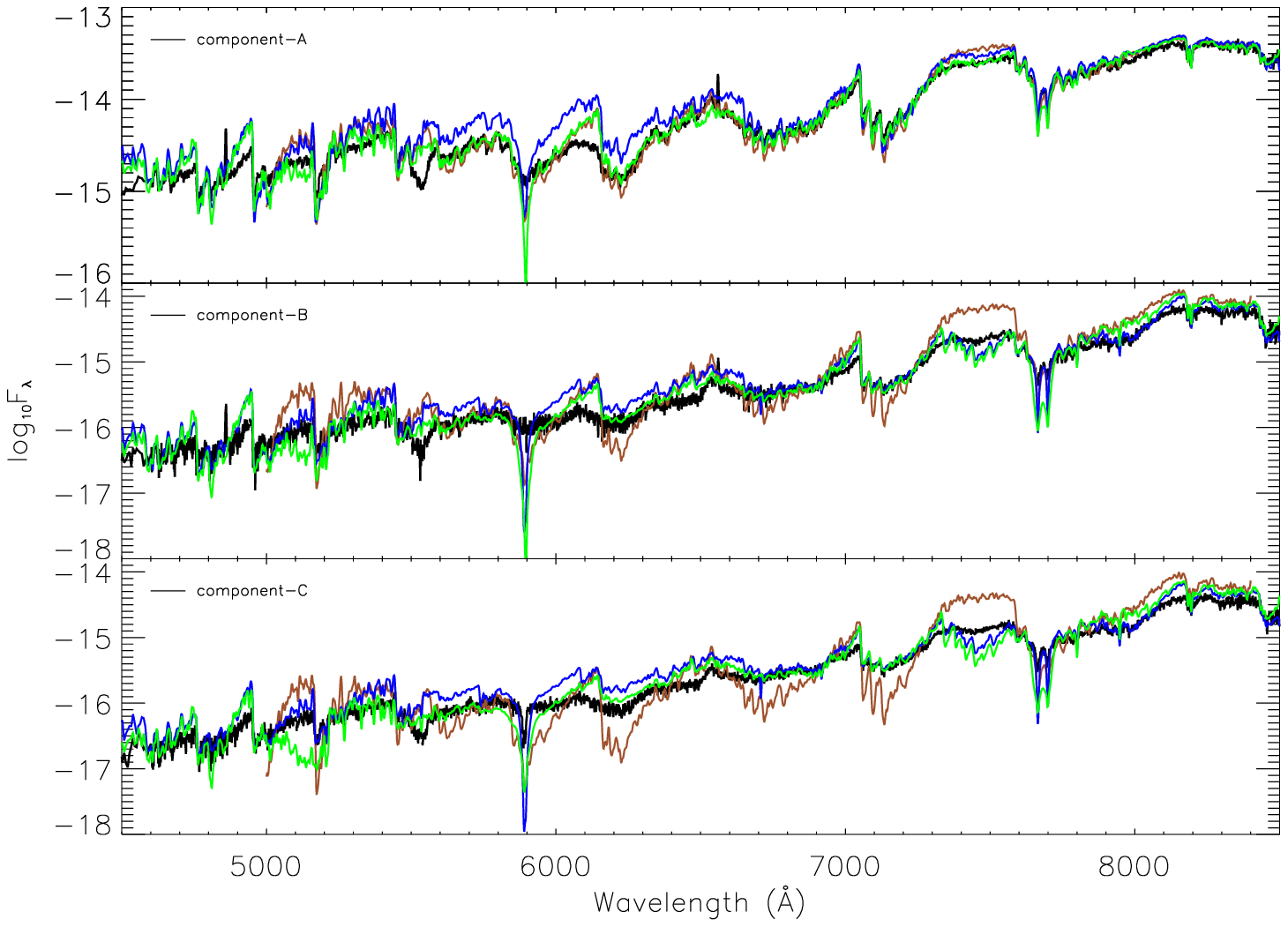}
\hspace{-1cm}
 \includegraphics[width=8.2cm]{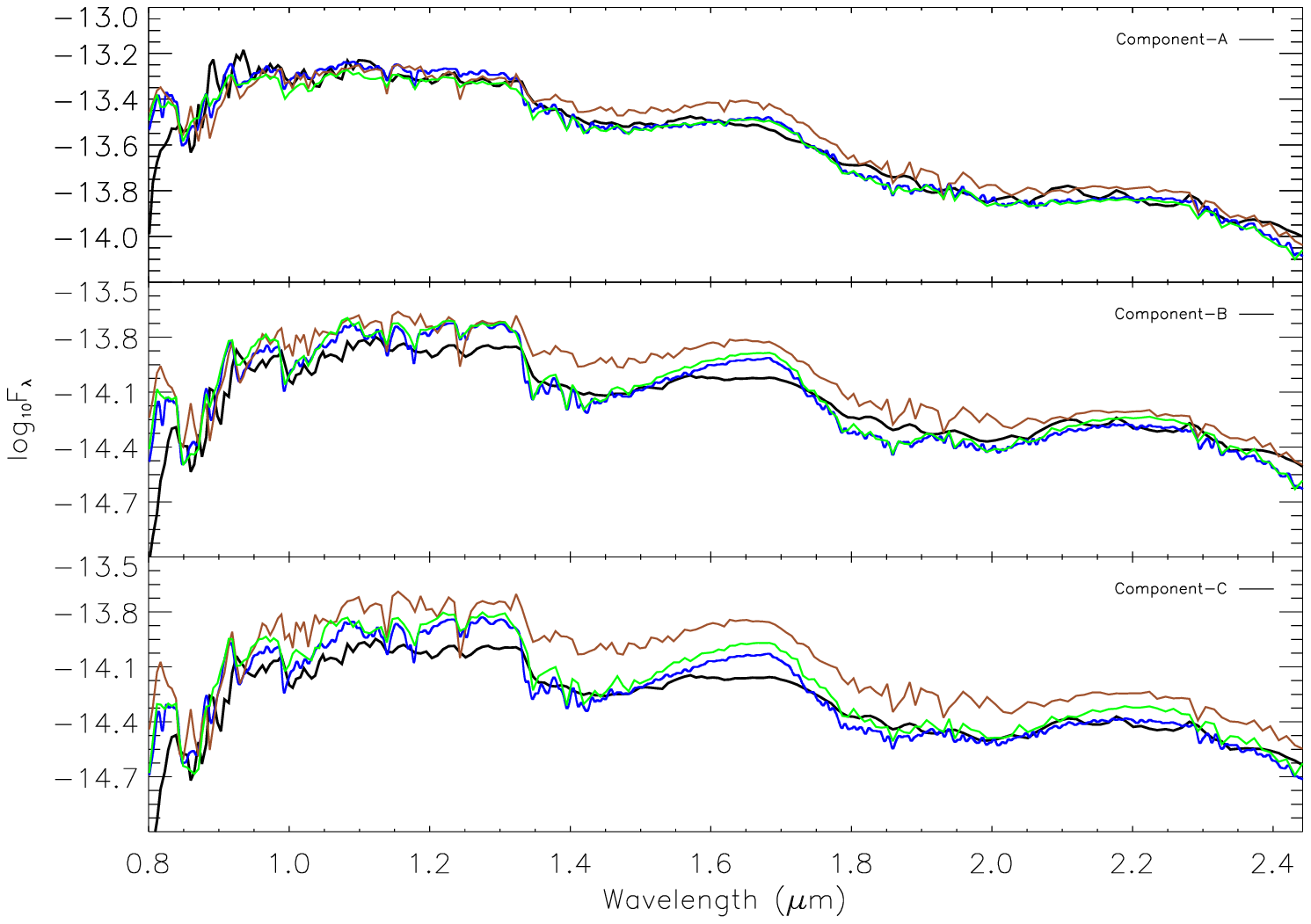}

  \caption{\footnotesize
     Fig. 8 (left) and 9 (right) of Rajpurohit et al. (2012). In each of the three panels, 
     the spectra of the resolved components of the LHS\,1070 triple system \citep{Leinert1994} 
     --- observed with the Faint Object Spectrograph (FOS) on the left \citep{Leinert2000} 
     and with NICMOS on HST on the right ---  are compared to models of various authors. Black: 
     observed spectra. Green: best ${\chi}^2$ fit BT-Settl AGSS model Allard et al. (2012).  
     Blue: DRIFT model \citep{Helling08a}. And brown: MARCS model \citep{MARCS2008}. The 
     obtained parameters are given in \cite{LHS1070_2012}, but do not change by more then one 
     to two sigma from earlier estimations \citep{Leinert2000} based on the AMES-Dusty models 
     by \cite{Allard01}, given the grid parameter spacing (100\,K). The B and C components are 
     cool enough (\teff\ $= 2500$ and $2400$\,K respectively) for their SED to be affected by 
     dust formation.}
  \label{f:allard_f2}
\end{figure*}

Important revisions have been made to the solar abundances based on radiation hydrodynamical 
simulations of the solar photosphere, and to  improvements in the detailed line profile analysis. Indeed, two 
separate groups using independent RHD and spectral synthesis codes \citep[Caffau et al. 2011]{Asplund09}\nocite{Caffau2011} 
obtain an oxygen reduction of  0.11\,--\,0.19~dex (up to 34\%)  
compared to the previously used abundances of \citet{GNS93}.  
Since the overall SED of late K dwarfs, M dwarfs, brown
dwarfs, and exoplanets is governed by oxygen compounds (TiO, VO in the
optical and water vapor and CO in the infrared), the elemental oxygen
abundance is of major importance.
Fig.~\ref{f:allard_f2} shows an example of these effects, where several models
are compared to the optical to infrared SED of the M5.5, M9.5, and L0 dwarfs 
of the LHS\,1070 system. The BT-Settl model by Allard et al. (2012)\nocite{Allard2012} 
is based on the \cite{Asplund09} solar abundance values, while DRIFT models
by \cite{Helling08a} use the \citet{GNS93} solar abundances, and the MARCS 
model by \citet{MARCS2008} uses the values of \citet{Grevesse2007}. 
Inspecting Fig.~\ref{f:allard_f2}, one can see that the MARCS model show a 
systematic near-infrared flux excess, compared both to observations and the other models, 
which is probably caused by  the much lower oxygen abundance values of \citet{Grevesse2007}.
The oxygen abundances sensitivity of TiO bands is expressed as a reduced line blanketing effect at longer
wavelengths, participating in the water vapor profile changes \citep{AllardTiOH2O2000}.

\begin{figure*}[ht!] 
      \hspace{-0.5cm}
      \includegraphics[width=7.7cm]{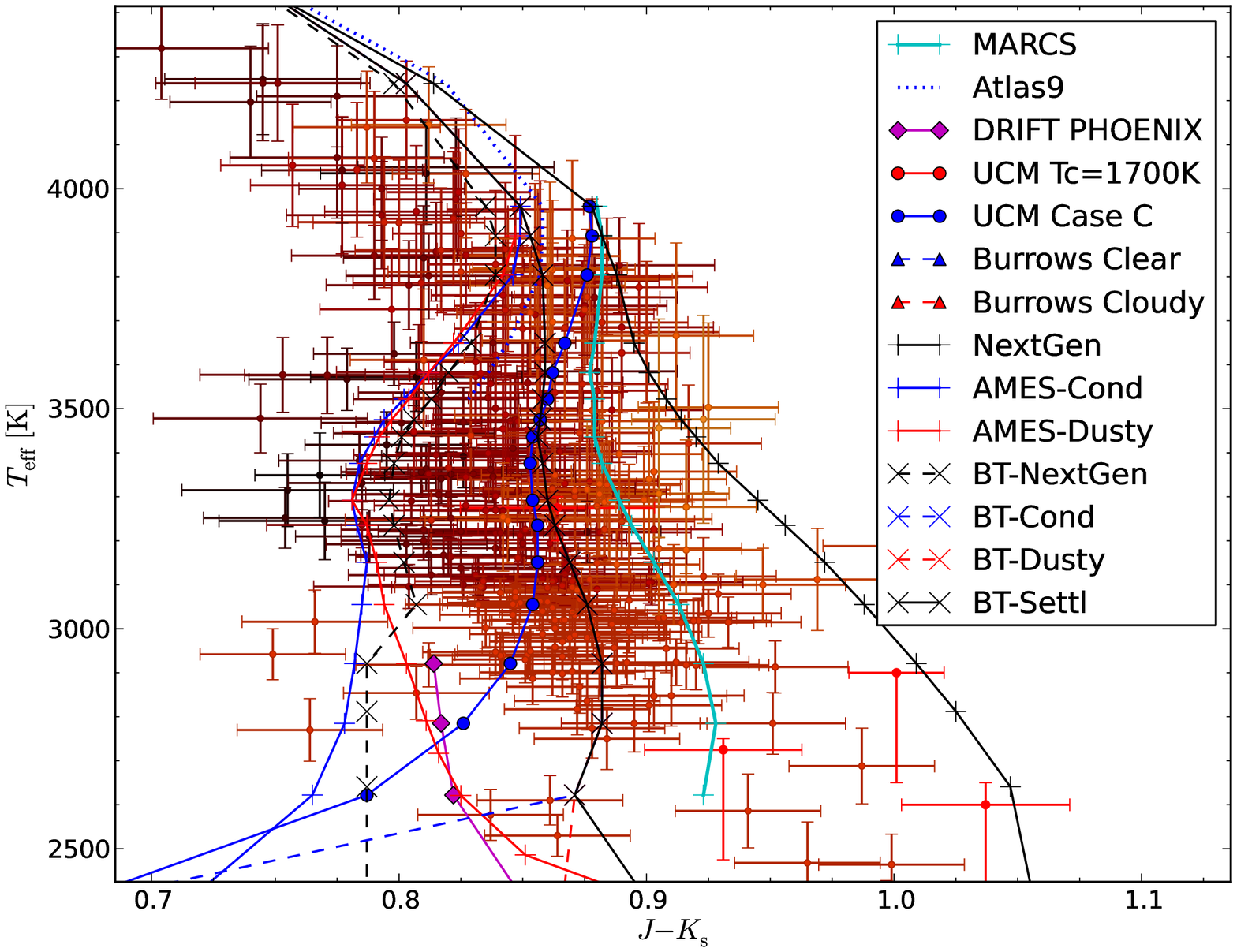}
      \hspace{-0.8cm}
      \includegraphics[width=7.7cm]{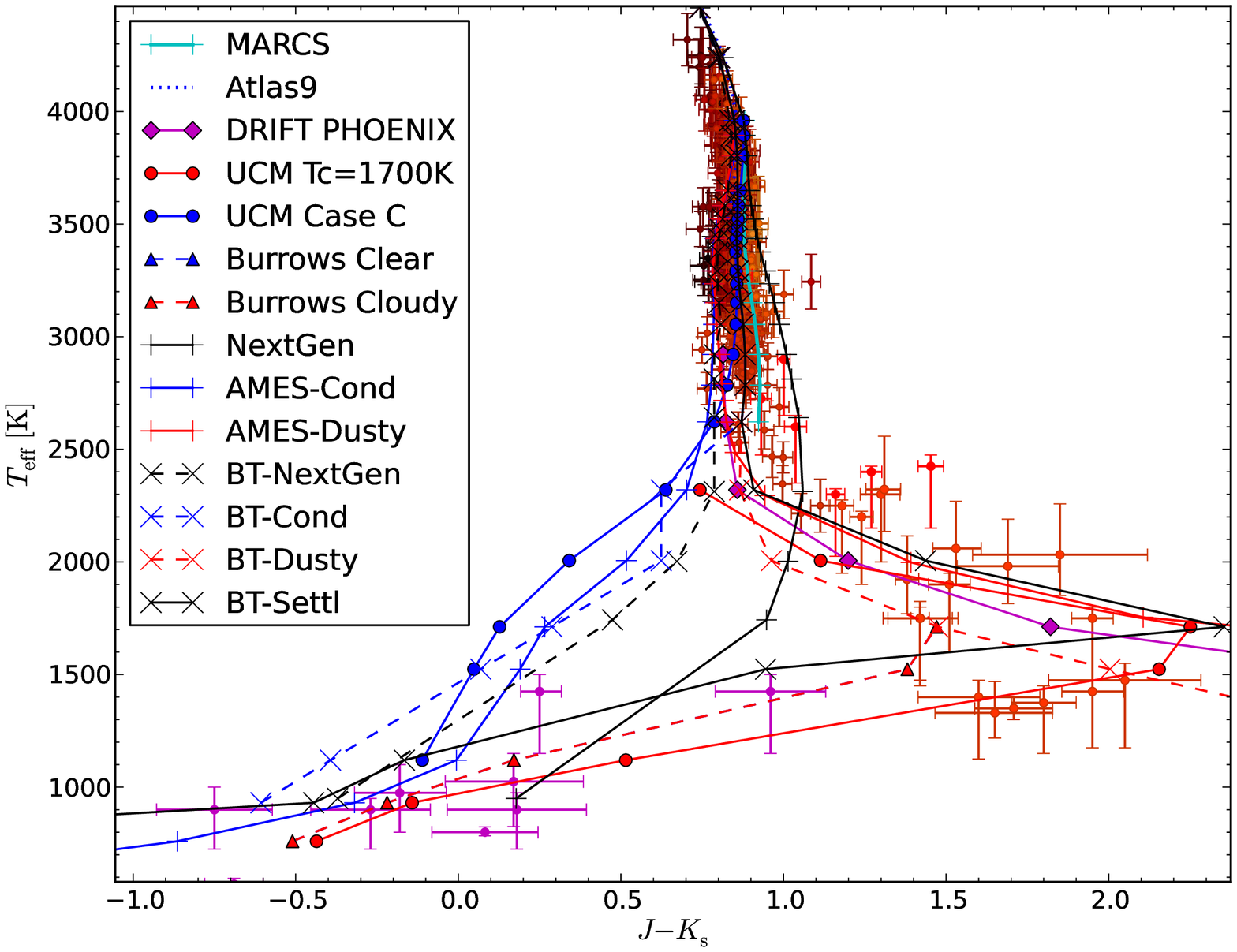}
  \caption{\footnotesize
     Estimated \teff\ and metallicity (lighter to darker tones) for 
     M dwarfs by \cite{MdwarfsTeff2008} on the left, and brown dwarfs by Golimowski et al. (2004) 
     and Vrba et al. (2004) on the right are compared to the NextGen 
     isochrones for 5 Gyrs  \cite{BCAH98} using model atmospheres by various authors:  
     MARCS by \cite{MARCS2008}, ATLAS9 by \cite{ATLAS92004}, DRIFT-PHOENIX by 
     \cite{Helling08a}, UCM by \cite{tsuji02}, Clear/Cloudy by \cite{Burrows06}, NextGen 
     by \cite{NGa}, AMES-Cond/Dusty by \cite{Allard01}, and the BT models by Allard et al. (2012).
     The region below 2900\,K is dominated by dust formation.  The dust free models occupy the 
     blue part of the diagram and only at best explain T dwarf colors, while the Dusty and DRIFT 
     models explain at best L dwarfs,  becoming only redder with decreasing {\teff}.  The BT-Settl, 
     Cloudy and UCM  $T_{\rm crit}=1700$\,K models describe a complete transition to the red in the 
     L dwarf regime before turning to the blue into the T dwarf regime.  The Cloudy model however 
     does not explain the reddest L dwarfs.
     } 
  \label{f:allard_f3a_Teff-J-Ks}
  \label{f:allard_f3b_Teff-J-Ks}
\end{figure*}

The influence of the solar oxygen abundance can also be clearly seen in 
Fig.~\ref{f:allard_f3a_Teff-J-Ks} which compares the \cite{MdwarfsTeff2008}  
\teff\ and metallicity estimates with the \cite{BCAH98} NextGen
isochrones (assuming an age of 5 Gyrs) using model atmospheres from various
authors.   The oxygen abundance effects are particularly highlighted by comparing the 
BT-Settl model based on the \cite{Asplund09} values with models based on earlier 
solar abundance values.  This is the case of the AMES-Cond/Dusty and BT-NextGen models
by Allard et al. (2001, 2012) which are based on the Grevesse et al. (1993)
solar abundances.  On can see that the higher oxygen abundance causes models to appear 
too blue by as much as 0.75 mag compared to models based on the \cite{Asplund09} values. 
The MARCS models  \citep{MARCS2008} based on the \cite{Grevesse2007} values show on 
the contrary a systematically increasing excess in $J-K_s$ with decreasing {\teff}.    The models 
are most sensitive on the solar oxygen abundances for M dwarfs around 3300\,K, i.e. 
at the onset of water vapor formation.

The NextGen model by \cite{NGa} dates too far back and suffers from too much 
opacity differences (incompleteness essentially) to participate in this illustration.  
In fact, this plot helps to conclude that using the NextGen models caused a systematic overestimation   
of {\teff} for VLM stars. It is interesting to note that all models appear too red in the K dwarf
range above 4000\,K. This may be due to an under representation of the K dwarfs in this diagram. 
The unified cloud model (hereafter UCM) by  \cite{tsuji02} show a completely different behavior 
in this diagram, sharing the colors of NextGen or even MARCS models at 4000\,K, but diverging 
towards the BT-Settl colors at  3500\,K to finally cross-over to bluer colors as dust begin to form 
and affect the SED below 2600\,K.

The various model atmospheres have not been used as surface boundary condition to interior and 
evolution calculations, and simply provide the synthetic color tables interpolated on the published 
theoretical isochrones \cite[]{BCAH98}.   Even if the atmospheres partly control the cooling and 
evolution of M dwarfs \cite[]{CB97}, differences introduced in the surface boundary conditions by 
changes in the model atmosphere composition have negligible effect. 

\subsection{Metallicity}

The ground work in understanding the metallicity effects on the SED and colors of VLM stars has been
established by \citet{AllardPhDT90} and \citet{AH95}, and summarized by \cite{ARAA97}. The main effects
of reducing the metallicity are the gradual disappearance of the double-metal molecules of importance 
for the overall opacities and the SED of VLM stars (TiO, VO, CO), and the increased pressure effects 
(atomic line widths, the strength of the hydride molecular bands including the well-known 
H$_2$ collision-induced absorption (CIA) bands in the $K$ bandpass) resulting from the increased transparency 
of the atmosphere.  This is illustrated by Fig.~5 of \citet{ARAA97}, which shows how these changes operate
in pushing the peak of the SED to the blue as metallicity decreases from solar to [M/H]=-2.0, and -4.0, and the 
optical is becoming brighter while the CIA opacities depress the $K$-band flux. 

The main difficulty in parametrizing M subdwarfs using pressure indicators are due to 
the fact that an atmospheric pressure increase, while obtained by decreasing the metallicity, 
can also be provided by increasing the surface gravity, and/or reducing the effective temperature.  
Moreover, subdwarfs have systematically higher gravities and smaller radii than solar type M dwarfs \citep{BCAH97}, 
compensating for the increased brightness of the more transparent atmosphere at optical wavelengths.  
Thus, disentangling these effects can be difficult to impossible using low spectral resolution or broad band 
colors,  and detailed high resolution studies are necessary.  Work is in progress using the zeta index (or TiO/CaH flux ratio) 
by Lepine \& Scholz (2008)\nocite{Lepine2008} and  SDSS g-r and r-i colors (B\'arbara Royas Ayala and 
S\'ebastien L\'epine, private comm.).

In the substellar regime, the metallicity effects are similar to those enumerated above, but adapted to their cooler 
composition, with the difference that the SED of T dwarfs is not expected to become bluer with decreasing metallicity 
as for M dwarfs. This is due to the already extreme transparency of their atmosphere (due to condensation), and the 
resulting extreme strength and width of optical alkali spectral doublets (in particular Na\, I\,D and K\,I at 0.77$\mu$m).

Even though some metal-poor L dwarfs have been identified by Burgasser et al. (2003, 2004, 2006)\nocite{Burgasser2003b,Burgasser2004,Burgasser2006} 
and \cite{Lepine2009}, metal-poor brown dwarfs have all chances to be T dwarfs by the age of the thick galactic disk (Population II) 
according  to evolutionary models. For this reason, and because T dwarfs are more readily modeled than the dustier L dwarfs 
(see section~\ref{s:Cloud} below), several studies such as those by \cite{Liu2007}, \cite{Leggett2010}, and \cite{Pinfield2012} have been done, 
despite the extreme difficulty tied to the poor spectral resolution of the observations and the fact that the SED of T dwarfs changes
very slightly with the atmospheric parameters compared to those of L and M dwarfs.  The submetallicity of L and T brown dwarfs is not 
extreme ([Fe/H] $> -0.5$). More metal-poor brown dwarfs would more likely be Y dwarfs.

\subsection{Disequilibrium chemistry}\label{NLCE}

Mixing has been held responsible, beyond its role in updrafting
condensible gas into the cloud forming layers above, for deviations
from gas phase chemical equilibrium in the atmospheres of ultracool 
T dwarfs, exoplanets, and Jupiter. 
These effects on their SED have been observed as an
excess of carbon monoxide absorption \citep[Griffith \& Yelle
1999]{noll97Gl229B}\nocite{gyCO99}. Similarly, ammonia has been shown
to be under-abundant  \citep{didierGl570}. More recently, carbon
dioxide, which under chemical equilibrium conditions is expected to
form in only very small quantities in hydrogen-rich atmospheres, has
been directly detected spectroscopically \citep{TsujiAKARI} and
inferred from mid-infrared photometry \citep{Burningham2011}. 

This is understood as the result of the fact that the formation of
methane and ammonia, which are expected to dominate the carbon and
nitrogen chemistries, respectively, in the low temperature limit under
equilibrium conditions \citep{LF06}, has to compete against upmixing from
the deeper and warmer atmospheric layers. If the local mixing rate is
high compared with the relevant reaction rates, their high-temperature
counterparts (CO, CO$_2$, N$_2$) can instead be observed in the upper
atmosphere in excess of their local chemical equilibrium concentrations. 
To estimate the formation timescales one needs to identify the most
efficient reaction path of formation and then isolate the
rate-limiting, i.\,e.\ slowest, step in this path. 
For the conversion of nitrogen to ammonia this net reaction and
limiting step is generally taken to be 
\begin{eqnarray}
  \label{eq:n2reac}
  3 \mathrm{N}_2 + 2 \mathrm{H}_2 = \mathrm{NH}_3 :\nonumber \\*
  \mathrm{N}_2 + \mathrm{H}_2 \rightleftharpoons 2 \mathrm{NH} 
\end{eqnarray}
according to \citet{Lewis1980}, with the resulting timescale given by 
\begin{equation}
  \label{eq:n2tau}
  t_{\mathrm{N}_2}^{-1} = 8.45 \times 10^{-8}
  e^{-8151/{T}} \mathrm{cm^3s}^{-1} [\mathrm{H}_2]
\end{equation}
where $T$ is the gas temperature in K and [H$_2$] the number density of
molecular hydrogen. Due to the strong temperature sensitivity this
makes nitrogen destruction quickly inefficient around 2000\,K, meaning
that the nitrogen-ammonia ratio in typical brown dwarf atmospheres is
fixed already in the deep convection zone. 

The case for the carbon monoxide to methane conversion is more
complex, where a variety of possible reaction paths and corresponding
timescales have been discussed in the literature. In an extensive
analysis of the reaction network \citet{Visscher2010} propose 
\begin{eqnarray}
  \label{eq:coreac}
  \mathrm{CO} + 3 \mathrm{H}_2 = \mathrm{CH}_4 + \mathrm{H_2O} :
  \qquad \nonumber \\
  \qquad \mathrm{H}_2 + \mathrm{CH_3O}  \rightleftharpoons 
 \mathrm{CH_3OH} + \mathrm{H} 
\end{eqnarray}
with the timescale given by 
\begin{equation}
  \label{eq:cotau}
  t_{\mathrm{CO}}^{-1} = k_\mathrm{R863}
  \frac{[\mathrm{H}_2][\mathrm{CH_3O}]}{[\mathrm{CO}]} 
\end{equation}
and estimate a reaction constant \\[0.3ex]
$k_\mathrm{R863} = 1.77 \times 10^{-22}\,T^{-3.09}e^{-3055/{T}}\,\mathrm{cm^3s}^{-1}$. 
The resulting timescales are at some variance with earlier estimates
\citep[Yung et al. 1988, Griffith \& Yelle 1999, and Schaefer \& Fegley 2010]{prinn77}\nocite{yung88,gyCO99,Schaefer2010}, 
but they are generally becoming longer than atmospheric mixing timescales only in the cooler
outer layers above the convection zone. This means that measurements
of the CO abundance are a sensitive probe of overshoot and other
mixing processes above the Schwarzschild boundary.




This mixing may be described by a single diffusion coefficient
introduced as an additional model parameter that can be inferred from
the height above which CO and CH$_4$ should be kept fixed at 
their relative abundances; however this method strictly only probes
mixing at this ``quench level'' \citep{SaumonMarley2008}. 
The RHD simulations of \cite{Freytag2010} have allowed to model the
underlying mixing processes as a function of height, although the
translation from the averaged hydrodynamic velocity field to the
molecular diffusion coefficient is still subject to some uncertainty:
the scaling behavior of stochastic waves (in stable regions of the
atmosphere, far away from the convection zones) is different from that
of developed turbulence (as found -- approximately -- in the deep
stellar convective layers or the thin cloud convection zone) or from
motions in overshoot layers.  
Fig.~13 of \cite{Freytag2010} shows how the former may be approximated
by scaling with some power of the Mach number of the flow.
Spectroscopic comparison with the strength of CO and CH$_4$ features,
e.\,g.\ in Figs.~25, 26 of \citet{King2010} demonstrates that these
latter chemical models with relatively fast rates are most consistent
with  observations.

\section{Cloud formation}\label{s:Cloud}

One of the most important challenges in modeling these atmospheres
is the formation of clouds. \cite{Tsuji96a} had identified
dust formation by recognizing the condensation temperatures of hot 
dust grains (enstatite, forsterite, corundum: MgSiO$_3$, Mg$_2$SiO$_4$, 
and Al$_2$O$_3$ crystals) to occur in the line-forming layers 
($\tau \approx 10^{-4} - 10^{-2}$)  of their models. The onset of this phase 
transition occurs in M dwarfs below \teff\ $= 3000$\,K, but the cloud layers
are too sparse and optically thin to affect the SED above \teff\ $= 2600$\,K.
The cloud composition, according to equilibrium chemistry, 
is going from zirconium oxide (ZrO$_2$), refractory ceramics 
(perovskite and corundum; CaTiO$_3$, Al$_2$O$_3$), silicates 
(e.g. forsterite; Mg$_2$SiO$_4$), to salts (CsCl, RbCl, NaCl), and 
finally to ices (H$_2$O, NH$_3$, NH$_4$SH) as brown dwarfs cool 
down over time from M through L, T, and Y spectral types \cite[Fergley \& Lodders 2006]{Allard01}\nocite{LF06}.
This crystal formation causes the weakening and vanishing of TiO and VO 
molecular bands (via CaTiO$_3$, TiO$_2$, and VO$_2$ grains) from the 
optical spectra of late M and L dwarfs, revealing CrH and FeH bands 
otherwise hidden by the molecular pseudo-continuum, and the resonance 
doublets of alkali transitions which are only condensing onto salts in late-T 
dwarfs.  The scattering effects of this fine dust is Rayleigh scattering which 
provides veiling to the optical SED, while the greenhouse effect due to the 
dust cloud causes their infrared colors to become extremely red compared 
to those of hotter dwarfs.  The upper atmosphere, above the cloud layers, is 
depleted from condensible material and significantly cooled down by the 
reduced or missing pseudo-continuum opacities.

One common approach has been to explore the limiting properties of 
cloud formation. One limit is the case where sedimentation or gravitational settling 
is assumed to be fully efficient. This is the case of the Case B model of \cite{tsuji02}, 
the AMES-Cond model of Allard et al. (2001)\nocite{Allard01}, the Clear model of 
\cite{Marley2002}, and the Clear model of \cite{Burrows06}. The other limit is the 
case where gravitational settling is assumed inefficient and dust, often only forsterite, 
forms in equilibrium with the gas phase. This is the case of of the Case A model of 
\cite{tsuji02}, the AMES-Dusty models of \cite{Allard01}, the BT-Dusty models of 
Allard et al. (2012)\nocite{Allard2012},  the Dusty model of \cite{Marley2002}, and the Cloudy model of 
\cite{Burrows06}.  To these two limiting cases we can add a third case also explored 
by several, which is the case where condensation is not efficient and the phase transition 
does not take place.  This is the case of the NextGen models of \cite{NGa}, of the 
BT-NextGen models of Allard et al. (2012)\nocite{Allard2012}, and the Case B models 
of \cite[not shown]{tsuji02}. 

The purpose of a cloud model is to go beyond these limiting
cases and define the number density and size distribution of
condensates as a function of depth in the atmosphere, and as 
a function of the atmospheric parameters. The discovery of
dust clouds in M dwarfs and brown dwarfs has therefore triggered the
development of cloud models building up on pioneering work 
in the context of planetary atmospheres developed by
\cite{Lewis69}, \cite{Rossow78}, and \cite{Lunine89}.  The Lewis model
is an updraft model (considering that condensation occurs in a gas
bubbles advected from deeper layers). By lack of knowledge of
the velocity field and diffusion coefficient of condensates in the
atmospheres of the planets of the solar system, Lewis simply assumed
that the advection velocity is equal to the sedimentation velocity,
thereby preserving condensible material in the condensation
layers. This cloud model did not account for varying grain sizes 
(these naturally vary as a function of depth in the cloud layers). 
Rossow, on the other hand, developed characteristic timescales 
as a function of particle size for the main microphysical processes 
of importance (condensation, coagulation, coalescence, and 
sedimentation). The curve intersections 
gives an estimate of the condensate number densities and mean 
grain sizes. However, this model made several explicit assumptions 
concerning the efficiency of supersaturation, the coagulation, etc.  


\cite{Helling08b} have compared different cloud models and their impact 
on model atmospheres of M and brown dwarfs.  Most cloud models define 
the cloud base as the evaporation layer provided by equilibrium chemistry. 
In the unified cloud model of Tsuji et al. (2002, 2004)\nocite{tsuji02,tsujiCloudII} 
a parametrization of the radial location of the cloud top by way of an adjustable 
parameter $T_{\rm crit}$ was used. This choice permits to parametrize the cloud 
extension effects on the spectra of these objects without resolving the cloud model 
equations. In principle, this approach does not allow to reproduce the stellar-substellar 
transition with a unique value of $T_{\rm crit}$ since the cloud extension depends on 
{\teff}. Indeed, the transparent T dwarf atmospheres can only exist if the forsterite cloud 
layers retract below the line-forming regions in those atmospheres.

\cite{AM01} have solved the particle diffusion problem of
condensates assuming a parametrized sedimentation efficiency $f_{\rm
sed}$ (constant through the atmosphere) and a mixing assumed constant
and fixed to its maximum value (maximum of the inner convection
zone). \citet{Marley2002} and \citet{SaumonMarley2008} found that their 
so-called Cloudy models could not produce the M-L-T spectral transition with 
a single value of  $f_{\rm sed}$. This conclusion prompted them to propose a 
patchy cloud model \cite{Marley2010}. We have not been able to obtain these 
models for comparison in this paper.

\cite{Allard2003} and Allard et al. (2012)\nocite{Allard2012} have developed \texttt{PHOENIX} version 15.05 
using the index of refraction of 55 condensible species, and a slightly modified version of  the 
Rossow cloud model obtained by ignoring the coalescence and coagulation, and computing 
the supersaturation consistently.  They density and grain size distribution with depth in the atmosphere 
is obtained by comparing the timescales for nucleation, condensation, gravitational settling or sedimentation, 
and mixing derived from the Mixing Length Theory for the convective mixing in the convection zones, exponential 
overshoot according to \cite{Ludwig2002,Ludwig2006}, and from gravity waves according to \cite{Freytag2010}. 
The cloud model is solved layer by layer inside out (bottom's up) to account for the sequence of grain species 
formation as a function of cooling of the gas.  Among the most important species forming in the BT-Settl model 
are ZrO$_2$,  Al$_2$O$_3$, CaTiO$_3$, Ca$_2$Al$_2$SiO$_7$, MgAl$_2$O$_4$, Ti$_2$O$_3$, Ti$_4$O$_7$, 
Ca$_2$MgSi$_2$O$_7$, CaMgSi$_2$O$_6$, CaSiO$_3$, Fe, Mg$_2$SiO$_4$, MgSiO$_3$,  Ca$_2$SiO$_4$, 
MgTiO$_3$, MgTi$_2$O$_5$, Al$_2$Si$_2$O$_{13}$, VO, V$_2$O$_3$,  and Ni. At each step, the gas phase is 
adjusted for the depletion caused by grain formation and sedimentation.  The grain sizes (a unique maximum value 
per atmospheric layer) are determined by the comparison of the different timescales and thus varies with depth to 
reach a few times the interstellar values (used in the dusty limiting case models) at the cloud base for the effective 
temperatures discussed in this paper.  While the BT-Settl model assumes dirty spherical grains in the timescales 
equations to calculate the growth and settling of the grains, it only sums the opacity contributions of each species 
in each layer as for an ensemble of pure spherical grains. 

\cite{Helling08a} and \cite{Witte2009} modified the \texttt{PHOENIX} code 
to compute the DRIFT-PHOENIX models, considering the nucleation of only seven of the most 
important solids (TiO$_2$, Al$_2$O$_3$, Fe, SiO$_2$, MgO, MgSiO$_3$, Mg$_2$SiO$_4$) 
made of six different elements. The cloud model is based on resolving the moment equations 
for the dust density accounting for nucleation on seed particles and their subsequent growth or 
evaporation, solving from top to bottom of the atmosphere. This model assumes dirty grains mixed 
according to the composition of each atmospheric layer. It uses composite optical constants resulting 
in absorption and scattering properties of the grains that are therefore different than those of the 
BT-Settl models, possibly producing more opaque clouds. However, since the opacities are 
dominated by atomic and molecular opacities over most of the spectral distribution in this 
spectral type range, the impact of those differences are difficult to identify. The largest 
differences between the BT-Dusty, BT-Settl and DRIFT models are the differences in the 
local number density, the size of dust grains, as well as their mean composition, which are 
the direct results of the cloud model approach. The DRIFT model includes, similarly to the 
BT-Settl model, mixing by convection and overshooting by assuming an exponential decrease 
in mass exchange frequency in the radiative zone. But it neglects the contribution of the gravity 
waves included in the BT-Settl model. 
 
The models using the limiting cases of maximum dust content describe adequately 
(given the prevailing uncertainties) the infrared colors of L dwarfs. 
The cloud-free limiting case models, on the other hand, allow to reproduce to 
some degree the colors of T dwarfs.  But pure equilibrium chemistry models without 
parametrization of the cloud extension in the atmosphere cannot reproduce the observed behaviour
of  the M-L-T transition, the dusty models only becoming redder and dustier with decreasing {\teff}, while
dust-free models miss completely the reddening due to the dust greenhouse effects  in the L dwarf
regime. Fig.~\ref{f:allard_f3b_Teff-J-Ks} shows this situation compared with the effective temperatures 
estimates obtained by integration of the observed SED (Golimowski et al. 2004, Vrba et al 2004)\nocite{Golimowski04,Vrba04}.
One can see from Fig.~\ref{f:allard_f3b_Teff-J-Ks} that the late-type M and early-type L dwarfs 
behave as if dust is formed nearly in equilibrium with the gas phase with extremely red colors in 
some agreement with the AMES-Dusty models.  The BT-Settl models (full black line) reproduce the main 
sequence down to the L-type brown dwarf regime, before turning to the blue in the late-L and T dwarf 
regime as a result of the onset of methane formation in the $K_s$ bandpass. The BT-Settl models 
succeed as good as the limiting case AMES-Dusty (full red curve), BT-Dusty  (dashed red curve), and 
UCM $T_{\rm crit}=1700$\,K (full red with big dots curve) at explaining the reddest colors of L dwarfs (assuming an 
age of 5 Gyrs).   The fact that a UCM model with $T_{\rm crit}$ value of 1700\,K succeeds rather well in reproducing 
the L-T transition suggest that the cloud extension is somewhat constant through that transition. The DRIFT 
models, on the other hand, (magenta with diamonds curve) reach slightly less to the red and do not extend low 
enough in temperature to explain the L-T transition.
The M-L transition is not reproduced by any of the different models, as shown by Fig.~\ref{f:allard_f2} where the 
CIFIST and BT-Settl models begin to show a $J$-band flux excess for the B and C components. This suggests that 
an additional element neglected thus far is at play, such as larger maybe porous grains. Indeed, all models assume thus 
far spherical  and  non-porous grains.  The choice of solar abundances and the completeness of the opacity databases 
used is also somewhat important. One sees in Fig.~\ref{f:allard_f3b_Teff-J-Ks} (on the right) that models based on the 
\cite{Asplund09} solar abundances reach to redder colors in better agreement with constraints above 2000\,K then other 
models.  The understanding of the M-L transition between \teff\ $= 2000$ and $2400$\,K is an extremely important 
regime for the study of extrasolar planets\ldots 

\section{Conclusions}

We have compared the behavior of the recently published model atmospheres from various authors across the M-L-T spectral 
transition from M dwarfs through L type and T type brown dwarfs and confronted them to constraints.  If the onset of dust formation
is occurring below \teff\ = 2900\,K, the greenhouse or line blanketing effects of dust cloud formation impact strongly ($J-K_s < 2.0$) 
the near-infrared SED of late-M and L-type atmospheres with $1300 <$ \teff\ $< 2600$\,K.  The BT-Settl models by Allard et al. (2012)\nocite{Allard2012} 
are the only models to span the entire regime. In the M dwarf range,  the results appear to favor the BT-Settl based on the \cite{Asplund09} 
solar abundances versus MARCS and ATLAS 9 models based on other values.  In the brown dwarf (and planetary) regime, on the other hand, 
the unified cloud model by \cite{tsuji02} succeeds extremely well in reproducing the constraints, while the BT-Settl models also show 
a plausible transition.  However, no models succeed in reproducing the M-L transition between 2900 and 2000\,K. This \teff\ range 
is similar to that of young (directly observable by imaging) and strongly irradiated planets (Hot Jupiters). 


\begin{acknowledgements}
The research leading to these results has received funding from the 
French ``Agence Nationale de la Recherche'' (ANR), the ``Programme 
National de Physique Stellaire'' (PNPS) of CNRS (INSU), and the
European Research Council under the European Community's Seventh 
Framework Programme (FP7/2007-2013 Grant Agreement no. 247060).
It was also conducted within the Lyon Institute of Origins under 
grant ANR-10-LABX-66. 


\end{acknowledgements}

\bibliography{allard_f}

\end{document}